\documentclass[aps, prl, showpacs, twocolumn, superscriptaddress]{revtex4} 
\usepackage{graphicx}
\usepackage{dcolumn}
\usepackage{bm}
\begin{document}
\preprint{PRL}
\title{Observation of a New Fluxon Resonant Mechanism 
 in Annular Josephson Tunnel Structures}%
\author{ C. Nappi}
\affiliation{%
Istituto di Cibernetica "E. Caianiello" del C.N.R. 
, I-80078 Pozzuoli, Italy 
}%
%
\author{M. P. Lisitskiy}
\affiliation{%
Istituto di Cibernetica "E. Caianiello" del C.N.R. 
, I-80078 Pozzuoli, Italy 
}%
\author{G. Rotoli}
\affiliation{INFM Coherentia, Istituto Nazionale di Fisica della Materia, Unit{\'a} di Napoli}
\affiliation{Dipartimento di Energetica, Universit{\' a}  di L'Aquila, Localit{\'a} Monteluco, I-67040 L'Aquila, Italy}
\author{R. Cristiano}
\affiliation{%
Istituto di Cibernetica "E. Caianiello" del C.N.R. 
, I-80078 Pozzuoli, Italy 
}%
\author{A. Barone}
\affiliation{INFM Coherentia, Istituto Nazionale di Fisica della Materia, Unit{\'a} di Napoli}
\affiliation{Dipartimento di Scienze Fisiche Universit{\' a} di Napoli "Federico II", I-80126, Napoli, Italy}
\date{\today}
\begin{abstract}
A novel dynamical state has been observed 
in the dynamics of a perdurbed sine-Gordon system. 
This resonant state, has been experimentally observed as a singularity in the
dc current voltage characteristic of an annular Josephson tunnel junction,
excited in the presence of a magnetic field. With this respect, it can be 
assimilated to self-resonances known 
as Fiske steps. Differently from these, however, we demonstrate, 
on the basis of numerical simulations, that its detailed dynamics involves 
rotating fluxon pairs, a mechanism associated, so far, to 
self-resonances known as zero-field steps.
\end{abstract}
\maketitle

Singularities, or resonances, in the current - voltage characteristic of a Josephson 
tunnel junction,
reflect the underlying dynamics of the phase difference between the two
superconducting junction electrode order parameters. Two kinds of singularities are usually 
observed: 
resonances, known as Fiske steps (FS), that arise when a magnetic field is externally 
applied in the junction plane and  the so-called zero 
field steps (ZFS), excited even in the absence of an external magnetic 
field, exclusively in large junctions (a Josephson junction is said to be large 
when one of its dimensions is wider than the Josephson penetration depth 
$\lambda_J$ \cite{LJJ}).   
Two different resonance mechanisms have been proposed to explain the existence of these 
current-voltage singularities: interaction of cavity modes with
the ac Josephson effect and fluxon oscillations.
For small junctions, the theory of Fiske steps developed by Kulik \cite{Kulik}, 
based on the excitation of e.m. standing waves, accounts  for the 
experimental observations. In large junctions, the fluxon based picture, first 
proposed by Fulton and Dynes \cite{Fulton}, constitutes the convincing explanation frame of 
reference of all ZFS phenomenology. In the last approach, fluxons, or particle-like magnetic 
flux quanta (solitons), shuttle back and forth along the extended dimension of the junction.
 Here the
relevant equation is the one-dimensional sine-Gordon equation with appropriate perturbing 
terms and b.c. \cite{Parmentier}.
There have been attempts to extend the Kulik theory to long 
junctions such that a single type of analysis could work  for both FS and ZFS.  
Alternatively, the idea that fluxon propagation was responsible 
also for existence of FS, besides ZFS, in long junctions was put forward. This hypothesis, sometimes referred as 
"Samuelsen hypothesis" \cite{Ern,OlsenS}, is based on the simple observation that an applied magnetic field renders the junction dynamics asymmetric through the boundary conditions, 
i.e., the fluxon propagation becomes unidirectional: fluxons enter one of the junction edge and
annihilate on the opposite one.
The situation has been partially redefined by a 
number of experiments performed by Cirillo et al. \cite{Cirillo}. In these experiments 
there is evidence of a possibility which can account for both descriptions. 
There would exist separate 
regimes in long junctions, depending on the intensity of applied magnetic field, in which 
the two mechanisms are 
well separated and active. At small fields the fluxon picture applies, for larger fields, 
i.e. beyond the  threshold represented by $H_0 = 2\lambda_J j_c$ \cite{Thinkham}, 
where $j_c$ is 
the maximum critical Josephson current density, the field penetrates stably the junction which 
starts to behave like a short junction as far as its magnetic properties are concerned. 
In other words the cavity mode mechanism described by Kulik
operates beyond $H_0$ while the Samuelsen hypothesis acts at low field values, when the external  
magnetic field 
localizes at the edges of the junction.\\ 
Hereafter we report on observations made on an annular Josephson junction
, i.e., one in which the two electrodes are stacked superconducting  rings
coupled through a thin dielectric tunnel barrier.  
In annular Josephson junctions,  the above
 described phenomena
have a peculiar character due to the absence of fluxon reflections at the
boundaries. Much of the present knownledge about the classical fluxon
dynamics has been 
derived by experiments perfomed on annular Josephson junctions \cite{danesi,danesi2}.
Recently Kulik 
theory has been succefully extended to small annular junctions including 2d effects and 
the possibility of trapped fluxons \cite{CiLis,PRB2001}.  
Moreover quantum effects involving trapped fluxons and fluxon-antifluxon 
(F-AF) pair generation have been 
shown in these devices \cite{Wall,Fistul}, so that fluxon dynamics 
in annular junctions remains of topmost interest  in view of their possible 
use as quantum controllable devices \cite{vortexqbit}. \\
In long annular junctions FS and ZFS appear at the asymptotic voltage positions
$V_n=n{\bar c}\Phi_0/L$, where n is an integer number, 
 $L$ is the length of the circumference of the junction, 
${\bar c}$ 
is the velocity of light in the junction (the ultimate fluxon velocity) and $\Phi_0$ the flux 
quantum. ZFS dynamics involves free 
propagation of fluxons (antifluxons) around the circle. 
Unless fluxons are inserted into the junction \cite{danesi2,Ustinov}, thus changing the 
so-called winding number, ZFS correspond in fact to the propagation of fluxon-antifluxon 
pairs, with zero winding number. 
On the other hand the Samuelsen hyphothesis for FS in long annular junctions corresponds 
to the following  picture:
in the presence of an external magnetic field, fluxon-antifluxon pairs 
are enucleated and successively
annihilated in correspondence of two opposite points along a diameter normal to the 
magnetic field direction where the 
tangential component of magnetic field $H_t$ is maximum and minimum respectively \cite{Monaco}. 
These two separate mechanisms, in the absence of trapped fluxons, provide through the
above formula for $V_n$ the even
voltage positions  (n=2,4,...) for ZFS's 
 and all the  positions (n=1,2,...) for the case of FS's.\\ 
In this letter we report on experimental and theoretical study of phase dynamics underlying 
resonances appearing in the I-V characteristic of a moderately extended annular junction.
From our analysis a new picture of the Samuelsen mechanism appears which permits to identify 
for the first time the {\it hybrid} character of certain fluxon based resonant dynamical states.
The underlying dynamics of the observed states is one in which {\it both}  Samuelsen 
unidirectional mechanism and ZFS freely fluxon propagation coexist. 
From this point of view they  stand halfway between FS and ZFS. For larger fields this hybrid dynamics
disappears and we 
are just facing the Kulik cavity mode dynamics.\\
Experiments have been performed on a
$Nb_{bottom}(150nm)$ /$Al_{bottom}(25nm)$ /$Al_{2}O_{3}$/ $Al_{top}
(25nm)$/$ Nb_{top}(50nm) + Nb_{wiring}(530 nm)$ 
 annular Josephson tunnel junctions. Thick Al layers were used in order to reach the low value of 
quality factor, $Q$, of the junction cavity and, thus, to satisfy the low $Q$ condition 
of the Kulik theory.
As for the junction geometry, we have chosen island type 
configuration with narrow ($\approx$ 8 $\mu$ m ) wiring 
leads.
The sample photograph is shown in the insert 
of Fig. 1 (a). Internal and external junction diameters were 61 $\mu$m and 91 $\mu $ m, 
respectively, giving a mean radius of ${\bar R}=38$ $\mu $ m. 
Physical parameters of the junction were the following:
$j_c=54$ A/cm$^2$, $\lambda_J=45$ $\mu$m, $H_0=0.6$ Oe, ${\bar l}=2\pi{\bar r}=5.23$, ${\bar r}={\bar R}/\lambda_J$  .
The Josephson penetration depth was practically equal to the external junction radius
so that our sample can be considered as a slightly extended annular junction. 
We have measured the dependence of the amplitudes of 
resonance steps of the current-voltage characteristic as a function of parallel 
magnetic field, $H$. The amplitudes of four resonances were accessible to 
be registred \cite{Eucas2001}. 
The measurement details are described in \cite{PRB2001}.  
Fig. 1(a) shows the experimental parallel magnetic field dependence of the amplitude of the 
third resonance step (open circles). 
We used our model theory for small annular junctions 
\cite{PRB2001} as a tool to discriminate between features which 
do not match with 
the assumption of a cavity mode mechanism. The solid curve of Fig.1(a) is the 
theoretical dependence derived from this theory. The experiment shows clearly two 
peaks, or lobes, in the field range $0.46 Oe <\mid H\mid < 1.0 Oe$ which the theory
 does not predict at all. 
\begin{figure}
\includegraphics[width=9cm,height=12cm]{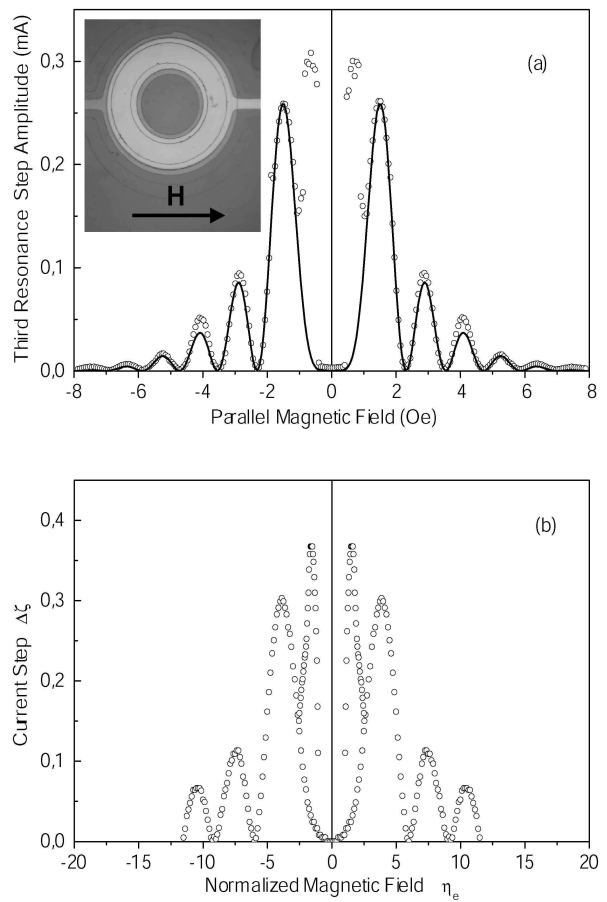}
\caption{\label{fig:epsart}a) Experimental magnetic field dependence of 
the current singularity  at voltage position corresponding to the third resonance step 
(open circles). 
The solid line has been obtained by an extension of Fiske step Kulik theory to 
two-dimensional 
annular junction case. Two  peaks are visible beside the Kulik predicted dependence of 
the FS. b) Simulated magnetic field dependence of the critical current of the third 
FS obtained by numerical solution of Eq.(\ref{eq:one}). In the simulation 
${\bar l}=5.23$, $\Delta r =0.33$, $\alpha=1/Q=0.15$.}
\end{figure}
This feature appears in a field region of the order of  $H_0$. According to the Samuelsen 
fluxon picture, 
three pairs should be 
enucleated at the point where
$H_t$ is maximum and annihilate on the opposite side. This would result 
in an increment of the phase of $6\pi$ in a period giving the required FS3 voltage.
However the dimensions of the junction assume here a critical relevance in that the actual
realization of the cited mechanism must cope with the number of fluxons and antifluxons which 
actually can accomodate into the junction. Differently from the case of one-polar states, i.e., 
only fluxons or only antifluxons trapped into the junction, existance of stable F-AF pairs 
requires that the junction be sufficiently extended to permit the pair dynamics. In particular
it appears problematic here, for ${\bar l}=2\pi{\bar r}<2\pi$, 
the occurrence of resonances involving more than two pairs \cite{Note1}.
So that the above described
Samuelsen mechanism 
requires some modifications induced by 
the junction having such a critical dimension.\\
\begin{figure}
\includegraphics[width=3.2in,height=6.0in]{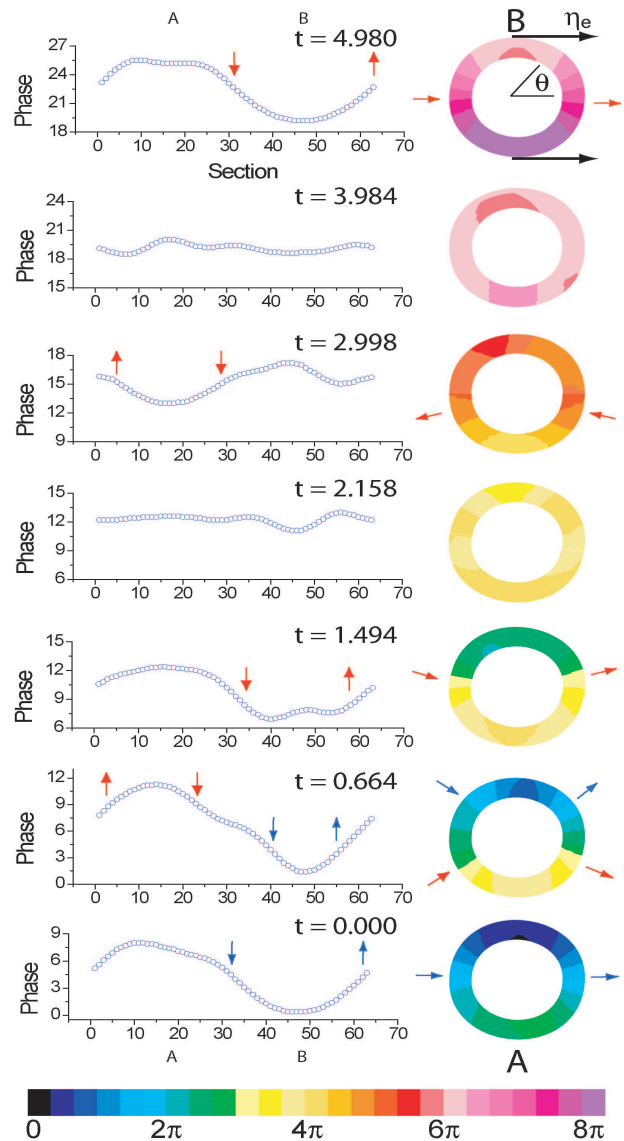}
\caption{(Color online) 
Time plots of the phase at the external border of the junction with the corresponding 2d phase
snapshots for the annular junctions considered in Fig. 1.$\zeta =1.104$ and $\eta_t=1.55$ just at 
the top of the lobe. Eq. (1) is discretized in 63$\times$10 spatial sections each long about 0.1$\lambda_J$. 
Section 63 is followed by section 1. The bias current drive is located just at sections 1-5 and 32-36. Point A and B are the maximum
tangential field locations where pair nucleation (annihilation) occurs. Blue arrows indicate the 
first F-AF pair, red arrows the second one. Directions of arrows indicate the fluxon polarity
}
\end{figure}
In order to investigate deeper this possibility, we solved numerically the following 2d perturbed 
sine-Gordon equation which governs the dynamics of the phase $\varphi(r,\vartheta, t)$ in an annular junction:
\begin{eqnarray}
\frac{1}{r^2}\frac{\partial^2 \varphi }{\partial\vartheta ^2}+
\frac{1}{r}\frac{\partial}{\partial r}\left( r\frac{\partial\varphi }{\partial r}\right)
-\frac{\partial^2 \varphi }{\partial t ^2}-\alpha \frac{\partial \varphi }
{\partial t}=\sin(\varphi )
\label{eq:one}
\end{eqnarray}
\begin{eqnarray}
\left.\frac{\partial\varphi }{\partial r}\right|_{r_e}=\zeta(\vartheta)-\eta_e \cos(\vartheta)\\
\left.\frac{\partial\varphi }{\partial r}\right|_{r_i}=-\eta_i \cos(\vartheta)\\
\varphi(\vartheta +2\pi)=\varphi (\vartheta )
\label{eq:bc}
\end{eqnarray}
In the above equation lenghts are normalized to the Josephson penetration depth
 $\lambda_J$ and times to
the Josephson plasma frequency $\omega_J={\bar c}/\lambda_J$. $\eta_e$,$\eta_i$ 
 are  the external and internal magnetic field normalized with respect 
to $\lambda_J j_c$ respectively. We assume a $\theta$ dependence of normalized bias current 
which is closer to the experiment, i.e., two thin leads feed the bias current to the junction.
In this case $\zeta=(j_b/j_c)\gamma$, with $j_b$ the bias current density and 
$\gamma=(\pi/\Delta\theta)(1+r_i/2r_e)\Delta r$, with $\Delta\theta$ the angular width of
current lead, $\Delta r$ the normalized junction width, $r_e$ ($r_i$)
the normalized external (internal) radius. 
Fig. 1b shows the result of simulation of the dependence 
of third resonance current on the external magnetic field $\eta_e$.
 The internal magnetic field was set to $0.85\eta_e$ to optimize
 the agreement with the experimental data, taking into account the screening effects. 
The simulation reproduces 
fairly well the result of Kulik theory for fields higher than $H_0$. The peaks
 at small fields are also well reproduced. From Fig. 1b it is seen that these 
lobes are coexisting with the beginning of the subsequent lobes as two separate branches. 
In fact the low field resonant peaks numerical solution branch
 is accessed by FS2 not from the McCumber branch \cite{LJJ}.\\   
The dynamics involved in the low field peaks is illustrated by the 
numerical simulation results reported in Fig. 2. The phase difference 
evolution between the two electrodes can be followed during a full 
period $T=3\Phi_0/{{V_3}{\omega _j}}\sim {\bar l}$. The phase distribution appearing 
in Fig.2 represents only a rough approximation to single fluxon
propagation, but the overall picture is sufficiently clear to draw the
following description.
A  F-AF pair (identified with the two points of steepest phase slope) 
is enucleated in the point A and it propagates up the positions 
indicated by the blue arrows as is seen at $t=0.000$ of Fig. 2. 
While it moves towards the opposite edge ($t=0.664$ in Fig. 2 blue arrows) 
a second pair enucleates in the point A and move towards the former 
(see again $t=0.664$ in Fig. 2, red arrows). 
After the first pair has reached the point B ($t=1.494$ in Fig. 2) it has an 
oscillation mode there because energy
is subtracted and the pair prepares to annihilate. Fig. 2 (red 
arrows) at $t=2.998$ shows that after the complex fourfold collision ($t=2.158$),
the second pair survives. In fact, it has still enough energy to turn back and complete 
the entire rotation at $t=3.984$, there the phase flatten and a new pair 
prepares to enucleate at point A (this last is shown at $t=4.980$). 
Things can be described as during the collision the second incoming pair gains energy at 
the expenses of the former oscillating pair, avoids annihilation, emerges 
from the collision and turns back. 
From the above analysis we see that the mechanism is hybrid: during a single period, it involves 
both a full rotating F-AF pair and a half propagating F-AF pair. 
From this point of view this dynamical state shares the nature of a Fiske resonance (the second Fiske resonance at $V_2$ in which the two pairs both annihilate) with that of a zero field resonance (the second zero field resonance at $V_4$ in which the two pairs both propagate 
around the junction making a complete turn). This shows that the fluxon picture in 
the presence of magnetic field is not limited to the half propagation mode, on the contrary 
it can be of complex (hybrid) nature. 
Numerical simulations show that hybrid dynamics on third resonance is stable in all range 
of normalized length ${\bar l}$ between about $4.5$ and $2\pi$.
The hybrid lobes for shorter length tend to become smaller with respect to Kulik subsequent 
lobes until they disappear when the junction length is less than about $4.5$. On the other 
hand for ${\bar l}>2\pi$ the junction dynamics switches to Samuelsen mechanism, i.e., 
three F-AF pairs are enucleated on one side and annihilated on the other one. 
In principle hybrid dynamics could exist whenever the length of the junction 
is not sufficient to accomodate the required number of pairs to achieve the correct 
voltage. The possibility of hybrid dynamics on other resonances, like fourth 
or fifth, is more questionable and maybe depends on the stability of these solutions against 
perturbations induced by other annihilating pairs on the ring.\\
In conclusion, we have observed a novel fluxon dynamical resonant state.  
In this both half propagating motion of a F-AF pair, due to external magnetic field, and 
propagation around the whole junction of an other F-AF pair coexist. 
This is probably a further striking observation of how nonlinear dynamics described 
by the sine-Gordon equation can be very complex even in a relatively simple structure such as 
an annular Josephson junction. It is easy to imagine how hybrid dynamics could be 
generalized to involve both asymmetric propagating and back and forth motion of fluxons also in 
standard long linear junctions. Beyond stability the ultimate existence of these solutions is tightly
related to the dimension of their attraction basin which can be difficult to access.\\
The possibility of ascribing the observed resonances to an hybrid complex dynamics requires 
a detailed analysis of the actual underlaying fluxon propagation mode. 
This should be taken into account specially in connection with the potential 
implementation of vortex qubit circuits based on multiply-connected Josephson 
structures \cite{vortexqbit}, where a severe control of all possible "states" is mandatory. \\
\begin{acknowledgments}
M.P.Lisitskiy is supported by MIUR under the project "Sviluppo di componentistica avanzata e 
sua applicazione a strumentazione biomedica". 
\end{acknowledgments}

\end{document}